\documentclass[american,farsi,english]{article}
\usepackage[LFE,LAE,T1]{fontenc}
\usepackage[utf8]{inputenc}
\usepackage{float}
\usepackage{textcomp}
\usepackage{graphicx}

\makeatletter
\usepackage{multicol}
\usepackage[farsi,english]{babel}

\makeatother

\usepackage{babel}
\DeclareTextSymbol{\guillemotright}{LFE}{62}
\DeclareTextSymbol{\guillemotleft}{LFE}{60}

\begin{document}
\selectlanguage{american}%

\title{How I Stopped Worrying about the Twitter Archive at the Library of
Congress and Learned to Build a Little One for Myself}

\author{Daniel Gayo-Avello\\
Department of Computer Science, University of Oviedo}
\maketitle
\begin{abstract}
Twitter is among the commonest sources of data employed in social
media research mainly because of its convenient APIs to collect tweets.
However, most researchers do not have access to the expensive Firehose
and Twitter Historical Archive, and they must rely on data collected
with free APIs whose representativeness has been questioned. In 2010
the Library of Congress announced an agreement with Twitter to provide
researchers access to the whole Twitter Archive. However, such a task
proved to be daunting and, at the moment of this writing, no researcher
has had the opportunity to access such materials. Still, there have
been experiences that proved that smaller searchable archives are
feasible and, therefore, amenable for academics to build with relatively
little resources. In this paper I describe my efforts to build one
of such archives, covering the first three years of Twitter (actually
from March 2006 to July 2009) and containing 1.48 billion tweets.
If you carefully follow my directions you may have your very own little
Twitter Historical Archive and you may forget about paying for historical
tweets. Please note that to achieve that you should be proficient
in some programming language, knowable about Twitter APIs, and have
some basic knowledge on ElasticSearch; moreover, you may very well
get disappointed by the quality of the contents of the final dataset.

\pagebreak{}
\end{abstract}

\section*{Twitter, light of my life, mire of my drive}

Twitter has become the \emph{de facto} source of data for most social
media research\footnote{If you are a regular to WWW, ICWSM, CHI, CIKM, ACL, HICSS, WSDM, EMNLP
or LREC you are painfully aware of that; if not please refer to \cite{rogers2013debanalizing}.
See also \cite{bruns2016twitter} for a rationale about preserving
Twitter as an important cultural artifact of our civilization.}; and that is not because of Twitter being the most popular online
social network or because of the high quality of the data it provides\footnote{For instance, user profiles at Twitter are extremely sketchy and basic
when compared to Facebook ones.}, but because it offers a convenient API to collect amounts of data
that seem–but rarely are–massive. 

Those researching Twitter in the broadest sense of the term tend to
worship–and hope to eventually reach–not one but two ``holy grails'',
namely, Twitter's Firehose (i.e., the whole stream of public tweets
published in real time) and Twitter's Historical Archive (i.e., the
whole set of public tweets since the beginning of the service in 2006).
Purportedly, such kind of data could provide extremely valuable insights
about our culture and society, in addition to allow a variety of natural
experiments about different kinds of social interactions–see, for
instance, \cite{watts2013computational}. 

The truth is that both sources of data are readily available, but
at high prices\footnote{I do not forget the Twitter Data Grants that allowed a limited number
of teams access to substantial amounts of Twitter data \cite{krikorian2014introducing};
however, I consider them a flash in the pan given that they have not
been offered anymore and, on top of that, only 6 out of 1,300 teams
(0.46\%) were awarded with one of them \cite{krikorian2014twitter}.} and, therefore, most researchers content themselves with less shiny–but
\emph{gratis}–materials such as the public streaming API (purportedly
a 1\% sample of the whole Firehose), and their own collections of
tweets–obtained either by filtering the streaming API or by using
the search API. 

Such kind of \emph{gratis} datasets face two major issues: on one
hand their representativeness is questioned (e.g., \cite{morstatter2014biased,tufekci2014big}),
and on the other hand they cannot be publicly released according to
Twitter's TOS (Terms of Service)\footnote{Certainly you can release lists of tweet ids but that means that other
researchers need to recollect the data again, and thus, it cannot
be properly considered as data sharing; still, it is the major if
not only approach to Twitter data sharing at this moment–e.g., \cite{mccreadie2012building}.}. This means that a huge amount of findings in the academic literature
cannot be replicated without enormous–and redundant–efforts, and they
may be perfectly wrong given that the data on which they rely is not
really representative of the whole of Twitter.

Given such state of matters, many of us welcomed the agreement between
the Library of Congress and Twitter to grant researchers access to
the Twitter Archive \cite{raymond2010tweet}. However, all that glitters
is not gold and the agreement had an important caveat: no substantial
portions of the archive could be available for downloading \cite{allen2013update}
what meant that researchers would need to physically access the archive
in order to perform any research. In addition to that, the amount
of tweets was so massive (170 billion tweets) that the Archive supposed
a huge technological challenge and real time queries were out of the
question\footnote{According to \cite{allen2013update} a single search could take up
to 24 hours to run. }. Hence, at the moment of this writing the Twitter Archive at the
Library of Congress is not available in any form and it seems that
it may remain that way \emph{sine die} \cite{mcgill2016can,scola2015library,zimmer2015twitter}. 

Still, there were other attempts to build searchable archives that
were contemporaries with LOC's announcement of the Twitter Archive
project, such as the now defunct Tweet Scan, or Summize that was eventually
acquired by Twitter and subsumed into \emph{search.twitter.com}. More
recently Jean Burgess provided a searchable archive of the first year
of Twitter \cite{burgess2013promised}, and, the fact is that Twitter
offers since 2014 full text search of the whole corpus of tweets \cite{zhuang2014building}
but no tools to allow the automatic downloading of the resulting tweets.

In other words, a searchable\footnote{I cannot but emphasize that the archive must be searchable, a simple
collection of tweets without any mean to perform queries is virtually
useless–and \emph{grep} does not count as a search tool.} Twitter Archive is technically feasible but there exists a tipping
point between the first year of Twitter and present day at which the
task is no more possible without an industrial-scale infrastructure–and
that alone explains the fees to access Firehose and historical tweets,
and also the failure to deliver of the LOC. 

Therefore, unlike Milo of Croton, we academics cannot hope to start
with a tiny archive and increase it slowly to finally index the whole
Twitterverse; instead, we must determine a final date–and thus a size–for
a feasible Twitter archive. Taking into account my resources (more
on this below) I chose as end date July 31, 2009 and the eventual
size of the dataset was 1,483,823,453 tweets\footnote{This makes my dataset comparable to the one described in \cite{cha2010measuring}
that is, needless to say, unavailable.}. The date was chosen to make this Twitter Archive contemporary with
the Twitter User Graph collected in \cite{kwak2010twitter} just if
my archive was some day released, even if by dumb luck. In the rest
of the paper I explain how you can build a similar archive while relying
on relatively little resources.

\section*{Gotta Catch'Em All}

\subsection*{A journey of a thousand miles begins with a single step }

To download historical tweets you just need to access the \emph{GET
statuses/lookup}\footnote{https://dev.twitter.com/rest/reference/get/statuses/lookup}\emph{
}API with a list of tweet ids; that endpoint allows you bulk downloading
up to one hundred tweets, and you can make a request every 5 seconds.
This means that you can download 1,728,000 tweets per day and, thus,
you would only need 850 days to collect the whole Twitter archive
described in this paper\footnote{Actually, that's the best case scenario in which you only request
those tweets that actually exist, if you were to try to download all
tweets between 0 and 3061013977 you would need 1,771 days. Later in
the paper I discuss how to reduce that time by exploiting a number
of curious facts regarding the tweet ids. }. Please note that such amount of time is well behind the 1,217 days
covered by the dataset; however, if you dare to try this exercise
you will find that by late 2009 the number of tweets is massive enough
to require more time to download than it took the world to tweet them. 

At this point, if you accept a task that is longer than a round-trip
mission to Mars you only have one problem: you need a list of tweet
ids. The good news is that during early Twitter ids are sequential,
the bad news is that current tweet ids are not, they are in no way
easy to ``guess'' \cite{king2010announcing}, and, hence, the approach
described in this paper does not work for tweets posted shortly after
mid-2010 \cite{singletary2010upcoming}\footnote{Anyway, by that date the number of tweets was about 170 billion and
it is doubtful you will be able to handle that amount of data–at least
with the resources available to the average academic.}. In other words, you may start producing tweet ids–one hundred at
a time– starting with 1 and finishing at 3061013977 (the midnight
between July 31 and August 1, 2009), submit them to the aforementioned
endpoint and relax. 

You surely have noticed that if ids were really sequentially produced
we should end much earlier–or find much more tweets; the truth is
that there are blank periods–i.e., ids that were not actually used–and
at times ids were increased not sequentially but in tens. Thus, to
avoid wasting resources trying to download nonexistent tweets you
may rely on the Octave code offered in the first of the appendices.

\subsection*{Desecrating the TOS}

Grace Hopper said, 
\begin{quotation}
``It's easier to ask forgiveness than it is to get permission.''
\end{quotation}
A friend of mine\footnote{Nudge nudge, wink wink.} applies that a
lot when interpreting Twitter's TOS; for instance, they looked for
a way to reduce those 850 days to a more amenable period of one month
and found that using 30 virtual machines could be a ``reasonable''
albeit non-kosher ``solution''. Such machines could be tiny with
barely the minimum RAM to run a Linux distro and a drive to store
a few gigabytes of data\footnote{On average a bulk of 5 million fully ``hydrated'' tweets takes 300
MB compressed and about 2.5 GB decompressed. } and, thus, they could easily fit in any reasonable academic cluster.
Each of them would require its own Twitter credentials which, in turn,
could require creating a bunch of Twitter accounts which, needless
to say, is far from being an orthodox reading of the TOS; actually
it is explicitly prohibited by Twitter's ``Automation rules and best
practices'':
\begin{quotation}
``Creating and/or automating serial (or multiple) accounts for overlapping
use cases is prohibited.''
\end{quotation}
When faced with such a fact my friend simply shouted ``Grace Hopper!''
while they plugged their ears and run in circles; they eventually
settled right down and reassured me that the key to not angering Twitter's
gods was avoiding throttling by respectfully waiting 5 seconds per
request. That way–they said–no user account or IP address would be
banned. It goes without saying that you should never ever violate
Twitter's TOS and, thus, if you follow the recommendations of my friend
do it at your own peril.

\section*{The Little Search Engine that Could}

Although I could understand my friend and their approach to use multiple
machines to parallelize tweet collection, the truth is that one of
my non-written goals was to provide a solution in a box. In other
words, the whole archive should fit in a single machine while at the
same time being searchable, if not in real time, at least in batch
mode. 

Taking into account what I've already mentioned in a footnote, a quick
back-of-the-envelope calculation would reveal that the whole dataset
should be about 88 GB compressed and around 750 GB when decompressed;
actually, it takes a little more than 140 GB compressed and almost
1 TB when decompressed.\textbf{ }That can barely be considered Big
Data\footnote{Unless you mean by Big Data ``something that doesn't fit into my
machine''.} but, still, storing such amount of data plus the index to search
it would require a muscle-workstation and not an average computer.
How muscled? I must confess I don't know because I was certain that
I was not able to afford it, much less to have a cluster to store
the archive \emph{sine die}. 

What I had instead was an Intel Core i7 with 8 GB RAM and a 512 GB
HDD, plus a budget of €400 to buy a SSD (Solid State Drive); thus,
the box was eventually indulged with a 1 TB Samsung SSD 850 EVO. The
question now was, what kind of data could be handled with those resources?
The truth is that quite a lot, provided that you are willing to make
some concessions. Indeed, the main first lesson to build a poor-man
searchable Twitter archive is that you must sacrifice most of Twitter's
metadata or, in other words, dehydrate the tweets a lot! What follows
is the ElasticSearch mapping that I eventually used:
\begin{verbatim}
{"twitter-archive": {
  "mappings": {
    "tweet": {
      "properties": {
        "created_at":{"type":"string"},
        "id_str":{"type":"string"},
        "in_reply_to_status_id_str":{"type":"string"},
        "in_reply_to_user_id_str":{"type":"string"},
        "lang":{"type":"string"},
        "text":{"type":"string"},
        "timestamp":{"type":"long"},
        "user_id_str":{"type":"string"}
      }
    }
  }
}
\end{verbatim}
As you may see\footnote{I'm assuming you have a good knowledge of the structure of a tweet;
should that not be the case please refer to https://dev.twitter.com/overview/api/tweets} I reduced the data to the barely minimum fields to work with tweets
while keeping information about replies and mentions, plus a timestamp
to allow range searches and chronological ordering of tweets\footnote{Please note that because of such decision the current version of the
archive does not support every kind of query; for instance, looking
for all the tweets produced or mentioned-at a given screen name is
not possible. In addition to that no information about the Twitter
user graph is stored.}. By doing that the size of the compressed dataset fell below 90 GB
and the indexed data was reduced to a little less than 400 GB; in
other words pretty ``maneuverable'' with limited resources.

Needless to say, even in that sparse shape it would have been quite
irresponsible to index all of the tweets within a single ElasticSearch
index and, actually, a relatively large number of them were created.
My recommendation is to store all of the 2006 data (181,217 tweets)
in one single index, use one index per month for 2007 and 2008 tweets
(an average of 11 million tweets per index), and one index per week
for tweets corresponding to 2009 (an average of 61 million tweets
per index).

Such kind of organization allows for range searches which do not require
all of the indices, and it still offers a reasonable throughput when
submitting queries to all of the indices. For instance, a convoluted
query such as
\begin{verbatim}
a OR about OR after OR all OR also OR an OR and OR any OR as OR at
OR back OR be OR because OR but OR by OR can OR come OR could OR 
day OR do OR even OR first OR for OR from OR get OR give OR go OR 
good OR have OR he OR her OR him OR his OR how OR I OR if OR in OR 
into OR it OR its OR just OR know OR like OR look OR make OR me OR 
most OR my OR new OR no OR not OR now OR of OR on OR one OR only OR 
or OR other OR our OR out OR over OR people OR say OR see OR she 
OR so OR some OR take OR than OR that OR the OR their OR them OR 
then OR there OR these OR they OR think OR this OR time OR to OR two 
OR up OR us OR use OR want OR way OR we OR well OR what OR when OR 
which OR who OR will OR with OR work OR would OR year OR you OR your
\end{verbatim}
can be served with the described infrastructure in about 2 minutes\footnote{If you wonder how much does it took to index the tweets I'm afraid
I cannot provide a solid figure since I did it in a number of batches;
however, it should be in the order of a fortnight.}; it is far from real time but it is much shorter than the hours queries
took in the LOC's Twitter Historical Archive \cite{allen2013update}.
For more ``reasonable'' queries such as \emph{obama}, \emph{``eating
a sandwich''} or \emph{``justin bieber''} the time required is
close to 5-10 seconds and well below 15 seconds–in other words, interactive
queries are feasible\footnote{You don't need to accept this at face value, please, try it by yourself
at \emph{http://danigayo.info/twitter-toddler/}.}. 

\pagebreak{}

\section*{They were the best of tweets, \protect \\
they were the worst of tweets}

On paper the dataset looks quite impressive: 1.48 billion tweets covering
the whole span from Twitter's creation on March 2006 to July 2009.
It reveals the evolution of the medium itself because it contains
the invention of hashtags, retweets and trending topics. Moreover,
it covers historically important events such as the 2008 US Presidential
Elections, the first Obama's inauguration speech or the 2009 Iran
Election protests. Finally, it does contain tweets in every major
language so it should be possible–at least in theory–to analyze international
events from different cultural perspectives.

In practice the data is much closer to anecdotal evidence on steroids,
and even though searches can be performed with any conceivable keyword
or set of keywords the fact is that we need much more than fancy time
series and tag clouds. Still, while we wait for a more thorough and
nuanced analysis methodology to exploit the dataset, we must content
ourselves with such superficial descriptions of the data.

The first thing we can explore is Twitter's growth; Fig. 1 reveals
that in little more than three years Twitter went from getting about
one thousand tweets per week to getting 100 million tweets per week;
in addition to that we may detect three different growth stages: the
early months when it was used as a private tool by its creators; the
initial public phase from July 2006 to early 2007; and finally the
continuous growing taking place after SXSWi 2007 \cite{douglas2007twitter}
when it went from one million tweets per week to 10 times that amount.
By projecting that trend we also find that the number of weekly tweets
would be close to 1 billion per week by mid-2010; taking that into
account it seems clear that to have a one-box archive it is very difficult
to go beyond late-2009/early-2010.

Also related to Twitter's growth and expansion we can check the evolution
of major languages in the platform–in per-mille tweets. Fig. 2 shows
their ratios from March 2007–the already mentioned major debout of
the service. Unsurprisingly, English is the dominating language in
the platform but languages such as Portuguese, Japanese and Spanish,
and to a lesser degree German and French have been substantially used
in the platform–from 10 to 100 tweets per thousand tweets. Still,
major global languages such as Arabic or Chinese are underrepresented
with about 1 tweet per thousand tweets.

\begin{figure}[h]
\includegraphics[bb=80bp 500bp 500bp 775bp,clip,width=12cm]{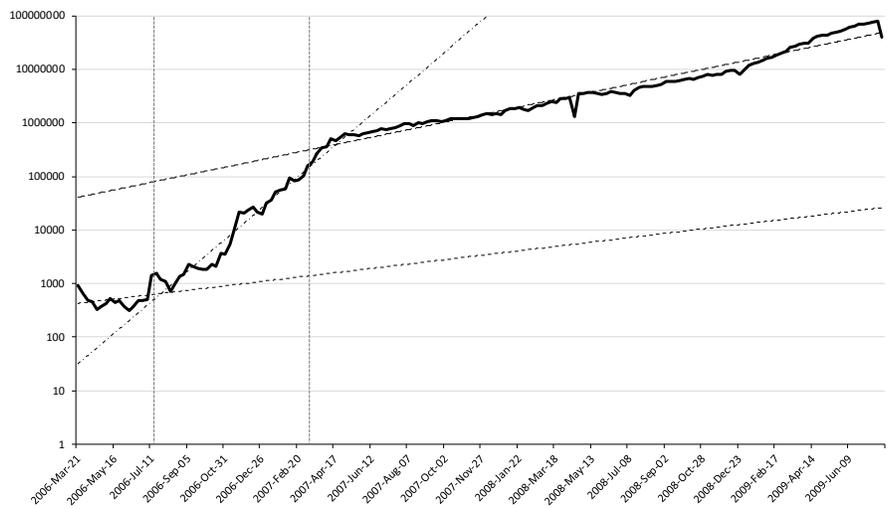}

\caption{\foreignlanguage{english}{Evolution (in logarithmic scale) of the raw volume of tweets. There
are three different stages in Twitter's growth: (1) from March 2006
to July 2006 when it is made public, (2) from July 2006 to March 2007
when it grows even faster, and (3) from March 2007 onward where the
growth is still exponential but at a slower rate. In three years Twitter
grew up from about 1000 tweets per week to 100 million tweets per
week (at the moment of this writing it is estimated to be in the order
of 500 million tweets per day!)}}
\end{figure}

\begin{figure}[h]
\includegraphics[bb=80bp 475bp 550bp 800bp,clip,width=12cm]{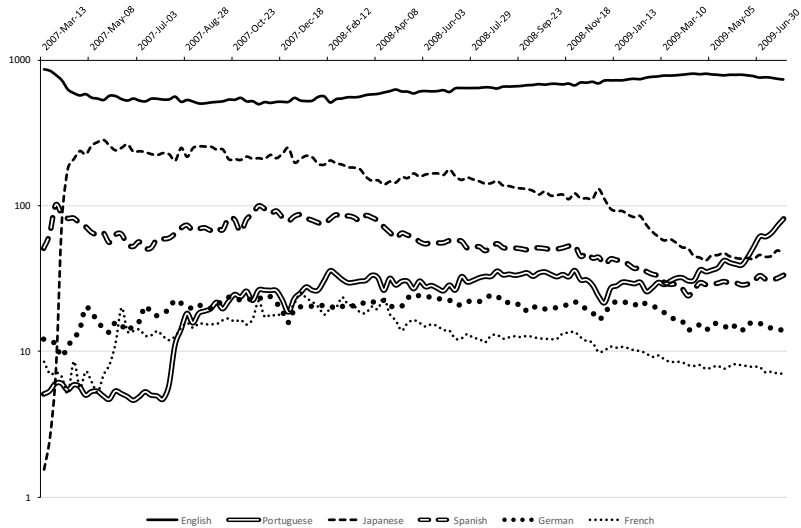}\caption{\foreignlanguage{english}{Major languages used in Twitter (in tweets per thousand tweets) after
the tipping point of SXSWi 2007. English is the dominant language
with about 900 out of 1000 tweets but Portuguese, Japanese and Spanish,
and to a lesser degree German and French were used from rather early
stages of Twitter. Other languages with presence (slightly below one
tweet per thousand tweets) are Chinese, Italian, Russian, Farsi, Turkish
and Arabic. It must be noted, however, that there is no correspondence
between number of speakers and volume of tweets; thus, Twitter was
at its beginnings a ``Western'' platform.}}
\end{figure}

Regarding the topics covered by Twitter users a much more thorough
analysis is required but we may dispell the popular trope of Twitter
being the place to say you are ``eating a sandwich''\footnote{Actually, the \emph{``eating a sandwich''} phrase dropped from 1‰
in mid-2006 to 0.004‰ in mid-2009}; as Fig. 3 reveals, \emph{``eating''} is not among the most frequent
actions appearing in English tweets. Moreover, there are a number
of actions–such as \emph{``reading''}, \emph{``looking''}, \emph{``watching''}
or \emph{``listening''}–that might suggest that Twitter was used
for live collective commenting of events from a very early stage. 

\begin{figure}[h]
\includegraphics[width=12cm]{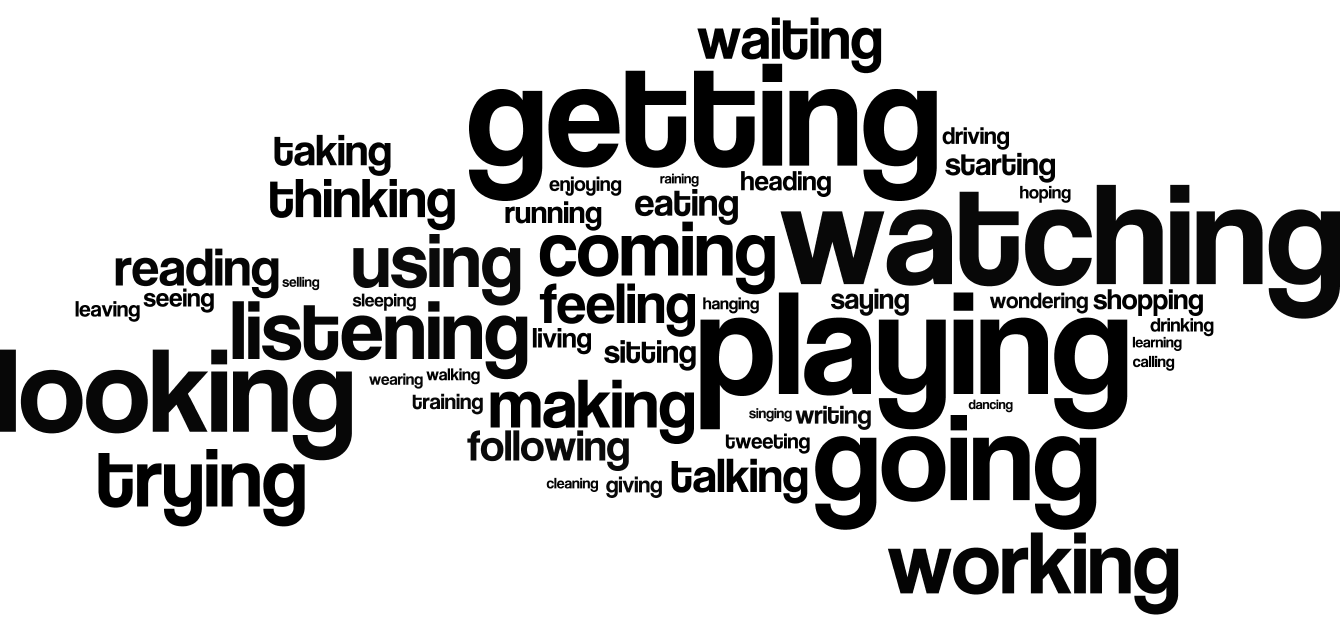}

\caption{\foreignlanguage{english}{A tag cloud showing the most frequent actions mentioned in Twitter.
Contrary to popular believe \emph{``eating''} is not the most prominent;
indeed, words such \emph{``watching''}, \emph{``looking''} or
\emph{``listening''} are the dominating ones. This would denote
that from a very early stage Twitter was used as a wall where live-comment
events the users were attending–either physically or in a mediated
way.}}
\end{figure}

In addition to that it seems that meta discussions–i.e., tweeting
about tweeting–are not the norm: they represent about 1‰ of the tweets.
Interestingly, \emph{``tweeting''} was not the word initially chosen
to mean ``posting a tweet'' but \emph{``twittering''}–see Fig.
4. 

\begin{figure}[h]
\includegraphics[bb=80bp 475bp 595bp 775bp,clip,width=12cm]{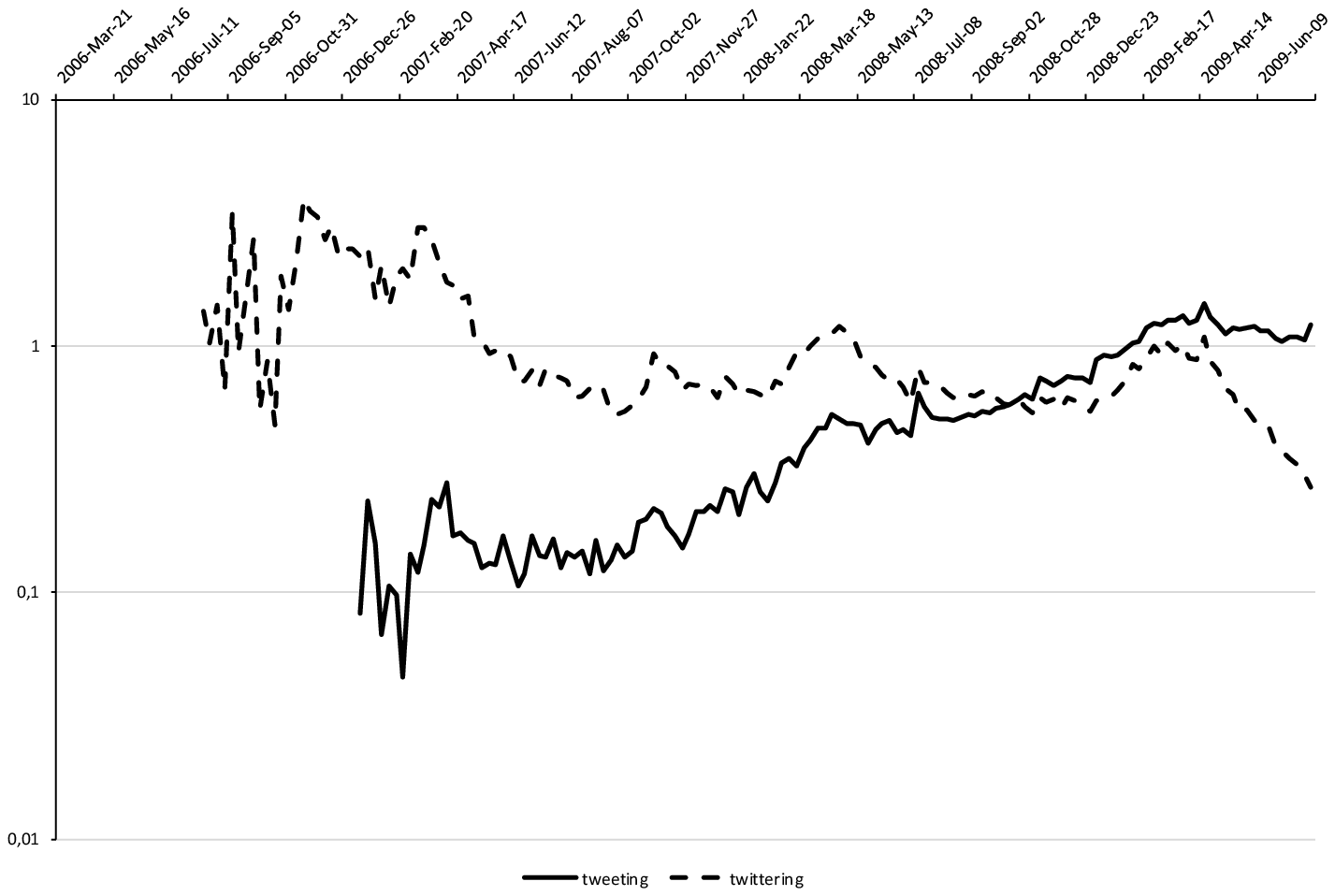}

\caption{\foreignlanguage{english}{Evolution of the use of \emph{``twittering'' }and \emph{``tweeting''}
per thousand tweets. Twittering was the preferred option to mean ``posting
a tweet'' during the first two years of the service although eventually
it has been replaced by \emph{``tweeting''}. Anyway meta tweets
are not the norm, a mere one per thousand tweets.}}
\end{figure}

Going on with Twitter-specific actions Fig. 5 shows the evolution
of hashtags, retweets, trending topics and follow Fridays. Please
note that those trends do not represent the actual volume of the corresponding
actions: For instance, \emph{``hashtags''} shows the use of the
keyword \emph{``hashtag{*}''} or the phrases \emph{``hash tag''
}and \emph{``hash tags''} but not the presence of actual hashtags\footnote{Should serve this as a warning to the reader, dehydrating too much
the tweets may eventually make some kind of queries virtually impossible;
in this case, I decided to drop the so-called entities–see\emph{ https://dev.twitter.com/overview/api/entities-in-twitter-objects}–from
the tweets and–by doing that–I lost the chance to search by hashtag
or URLs.}; similarly, the now common \emph{``rt'' }and \emph{``ff''} were
not used to point out retweets and follow Fridays because they had
a prior meaning, namely, the abbreviation of \emph{``right''} and
\emph{``Firefox''/``FriendFeed''}, respectively. Anyway, the graph
shows the approximate epochs for the invention of each of such behaviors.
Thus, the first mention of retweets (as retwitter) dates to February
2007–see Fig. 6; the invention of hashtags took place in August 2007–see
Fig. 7; and, finally, follow Fridays\footnote{The custom of suggesting users to follow every Friday.}
were suggested for the first time on January 2009–see Fig. 8.

\begin{figure}
\includegraphics[bb=80bp 475bp 595bp 775bp,clip,width=12cm]{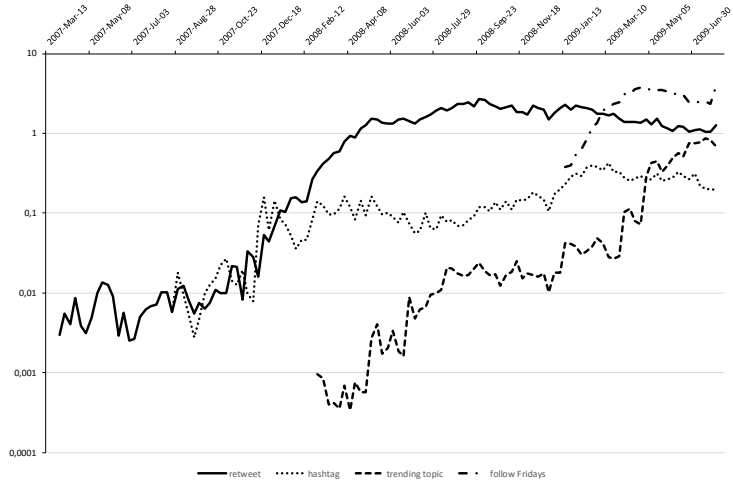}

\caption{\foreignlanguage{english}{Evolution of mentions about retweets, hashtags, trending topics and
follow Fridays–in per-mille tweets.}}
\end{figure}

\begin{figure}[h]
\centering{}\includegraphics[width=6cm]{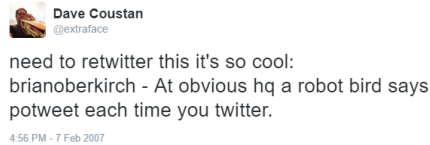}\caption{\foreignlanguage{english}{First mention to the retweet action (available at \emph{https://twitter.com/extraface/status/5361411}).}}
\end{figure}

\begin{figure}
\includegraphics[width=6cm]{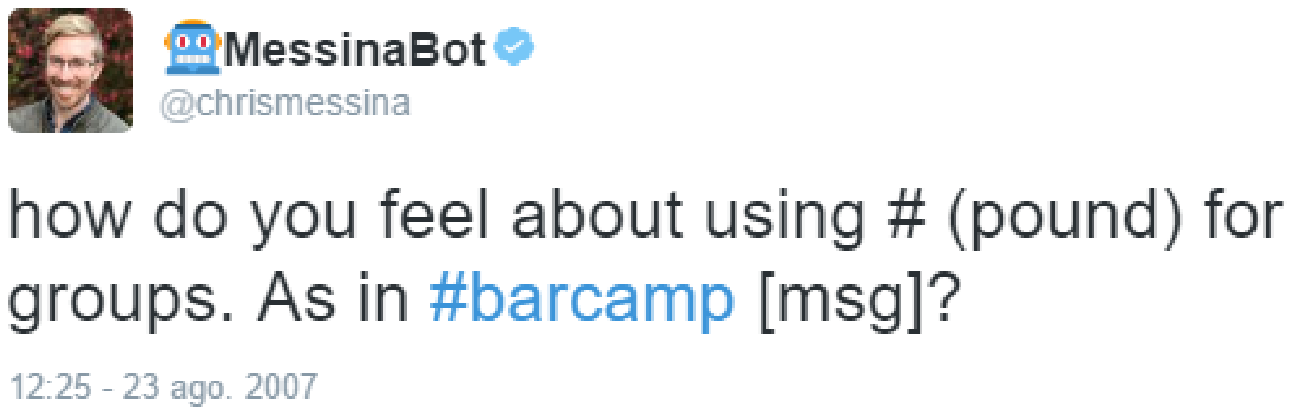}\includegraphics[width=6cm]{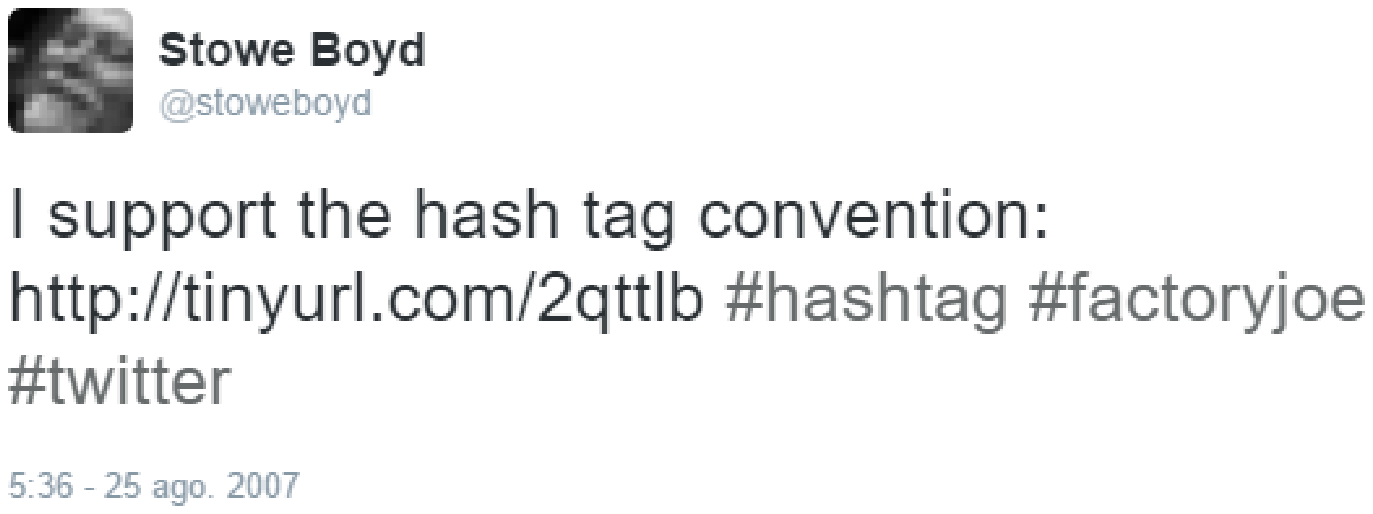}

\caption{\foreignlanguage{english}{The tweets inventing the hashtags and coining the neologism (available
at \emph{https://twitter.com/chrismessina/status/223115412} and \emph{https://twitter.com/stoweboyd/status/226570552},
respectively). Please note that the data associated with each profile
is the one at the moment of this writing and not the original ones.}}
\end{figure}

\begin{figure}
\centering{}\includegraphics[width=6cm]{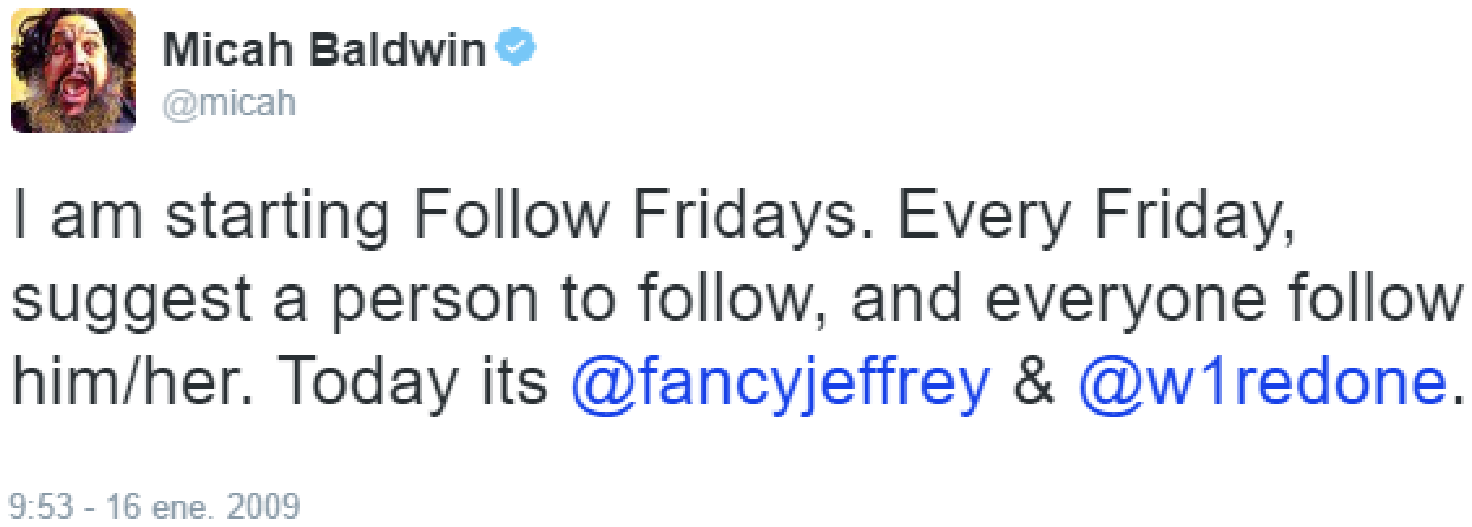}\caption{\foreignlanguage{english}{The first tweet suggesting Follow Fridays as we know them.}}
\end{figure}

The dataset also reveals that Twitter was used to live comment about
entertainment and shows very early. Fig. 9 shows the trends for just
three shows: American Idol, Hannah Montana and the Super Bowl; as
you may see the volume per thousand tweets increased with each season,
in fact–for the sake of visualization–the peaks of the Super Bowl
go off the chart, and already in 2009 tweets Super Bowl tweets amounted
to 11‰. 

The peaks of Hannah Montana reminds us that this dataset belongs to
a very different world; indeed it predates the age where an algorithm
was changed to avoid Justin Bieber constantly appearing in the list
of trending topics \cite{parr2010twitter}; in this dataset Justin
Bieber is not yet the most tweeted celebrity but just a YouTuber with
talent, actually, he appears in just 16,075 tweets while 285,614 tweets
mention Britney Spears.

\begin{figure}[h]
\begin{centering}
\includegraphics[bb=75bp 450bp 550bp 800bp,clip,width=12cm]{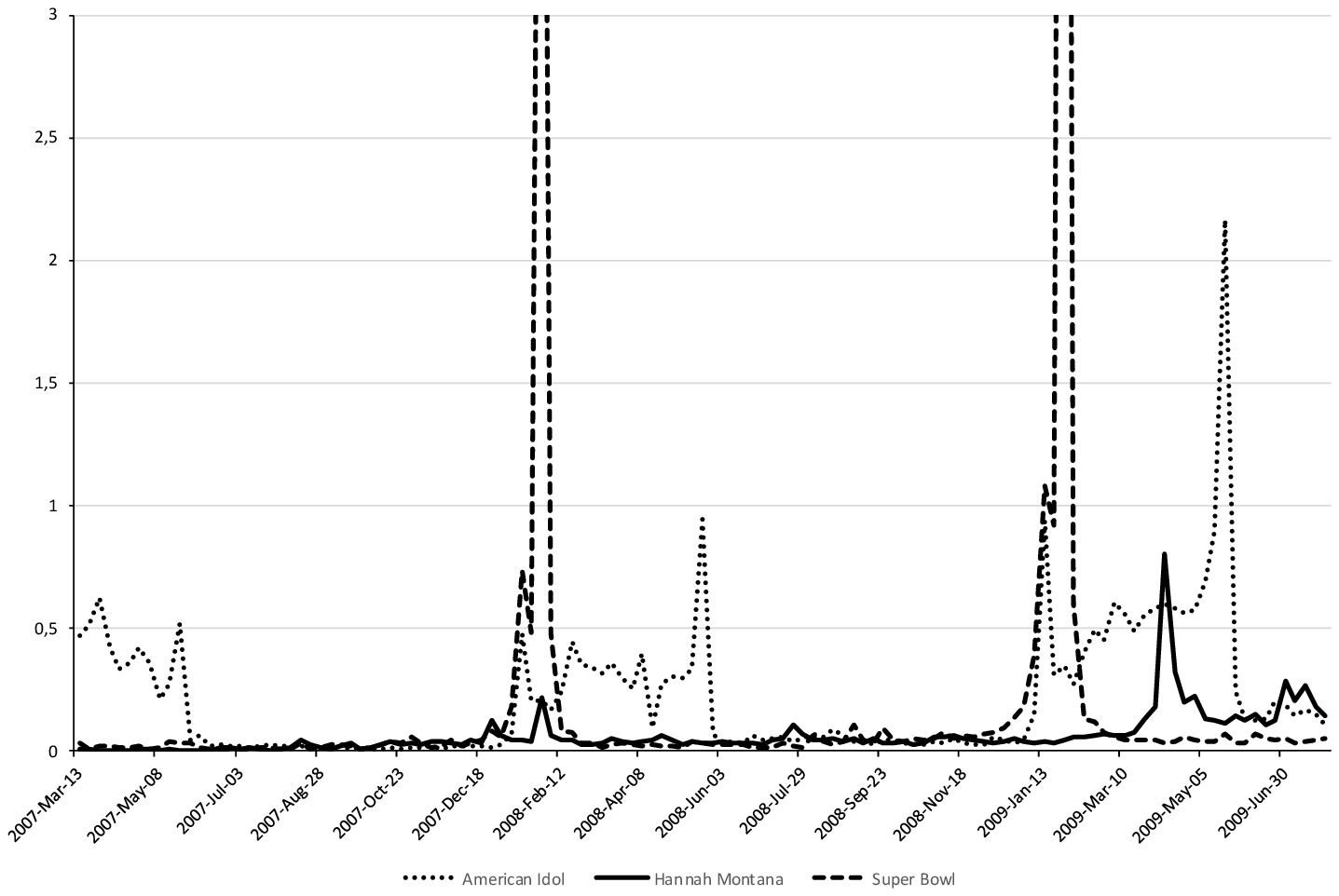}
\par\end{centering}
\caption{\foreignlanguage{english}{Twitter has been used from the beginning to live comment entertainment
events; such kind of application has increased with the years and
for some events like the Super Bowl it peaked at 11 tweet per-mille.}}
\end{figure}

Another revealing sign of the age of the dataset is found in the most
frequent URLs\footnote{The ratio of tweets containing URLs is relatively large: 25,3\% or
374,771,829 tweets.}: Up to 75\% of them belong to URL shortening and tracking systems,
being the most popular ones \emph{tinyurl.com}–42,87\% of the tweets–followed
by \emph{bit.ly}–29\% of the tweets, but we can also find \emph{is.gd},
\emph{twurl.nl}, \emph{ow.ly}, \emph{tr.im}, \emph{cli.gs}, \emph{snipr.com},
\emph{short.to}, \emph{snipurl.com}, \emph{migre.me}, \emph{mavrev.com},
\emph{tiny.cc}, \emph{snurl.com}, \emph{budurl.com}, \emph{post.ly},
or \emph{xrl.us}. Needless to say, most of such services are defunct
after Twitter providing their own URL shortener. 

Another service with a significant presence (7\% of the tweets) were
image hosting services such as \emph{twitpic.com}, \emph{yfrog.com},
\emph{flickr.com}, or \emph{mypict.me}; as with URL shortening, image
hosting was been eventually offered by Twitter and, hence, except
for those services predating Twitter, most of those in the described
dataset have already disappeared.

The third service which Twitter was lacking at those dates was geolocation
and there were a number of services to geolocate tweets such as \emph{myloc.me},
\emph{bkite.com}, or\emph{ loopt.us}; they supposed about 1\% of the
tweets those days but they are defunct now.

With regards to sites appearing on their own right and not because
of the need to provide lacking features to Twitter we can mention
\emph{digg.com}–0.31\% of the tweets, \emph{myspace.com}–0.24\% of
the tweets, \emph{youtube.com}–0.18\% of the tweets, and \emph{last.fm}–0.12\%
of the tweets. Except for MySpace which is virtually defunct, the
rest of sites are alive at the moment of this writing.

As a final note I would like to devote a few lines to cover two historical
events that were contemporary with the dataset and for which, purportedly,
Twitter would offer a privileged vantage point. They are the 2008
US Presidential Elections, and the 2009 Iranian Presidential Elections.

\begin{figure}[h]
\centering{}\includegraphics[bb=90bp 475bp 500bp 800bp,clip,width=12cm]{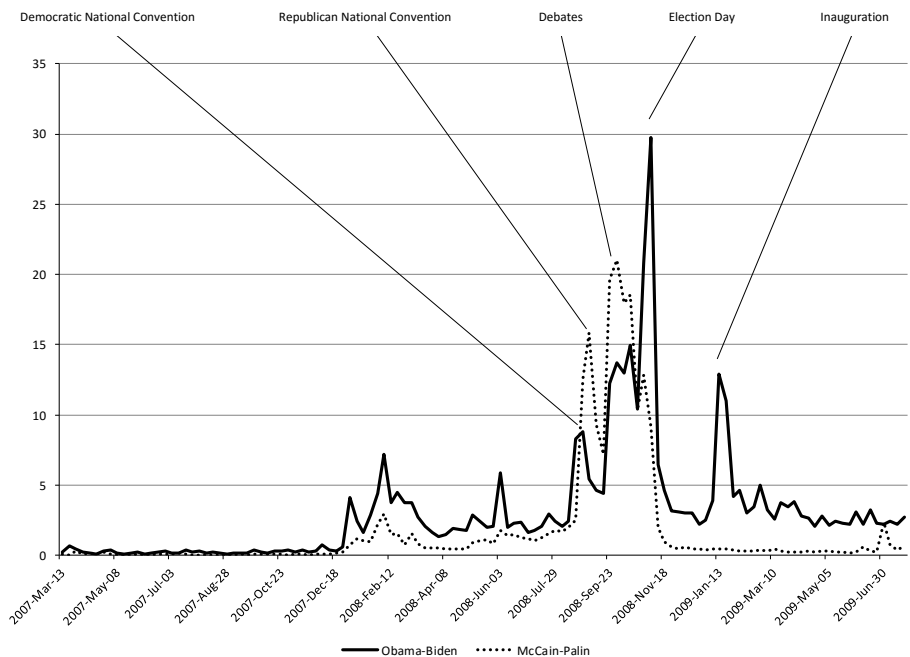}\caption{\foreignlanguage{english}{Evolution of tweets per-mille corresponding to the queries ``\emph{obama
OR biden'' }and \emph{``mccain OR palin''}.}}
\end{figure}

\begin{figure}[h]
\begin{centering}
\includegraphics[bb=80bp 675bp 520bp 775bp,clip,width=12cm]{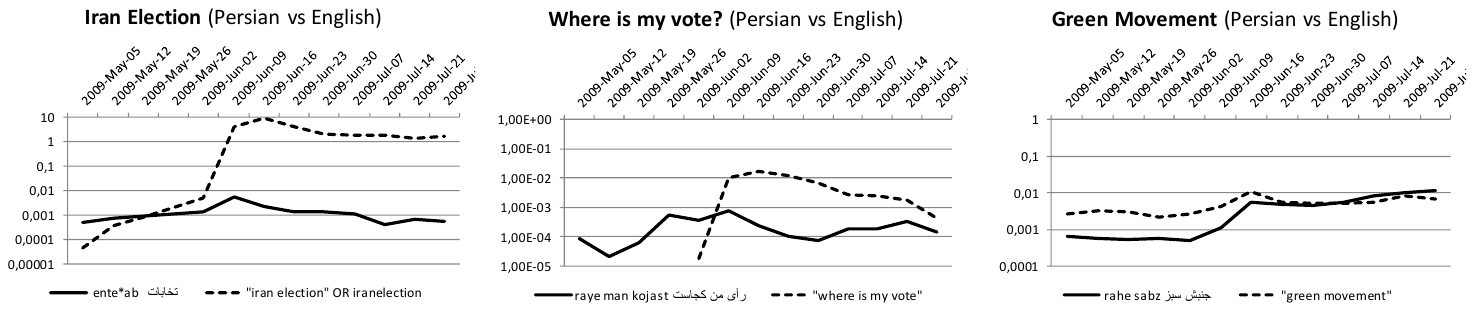}
\par\end{centering}
\caption{\foreignlanguage{english}{From left to right, evolution of the trends related to the topics
``Iran Election'', ``Where is my vote?'' and ``Green Movement''
in tweets per-mille in both English and Persian.}}
\end{figure}

\begin{figure}[h]
\begin{centering}
\includegraphics[bb=90bp 475bp 525bp 875bp,clip,width=12cm]{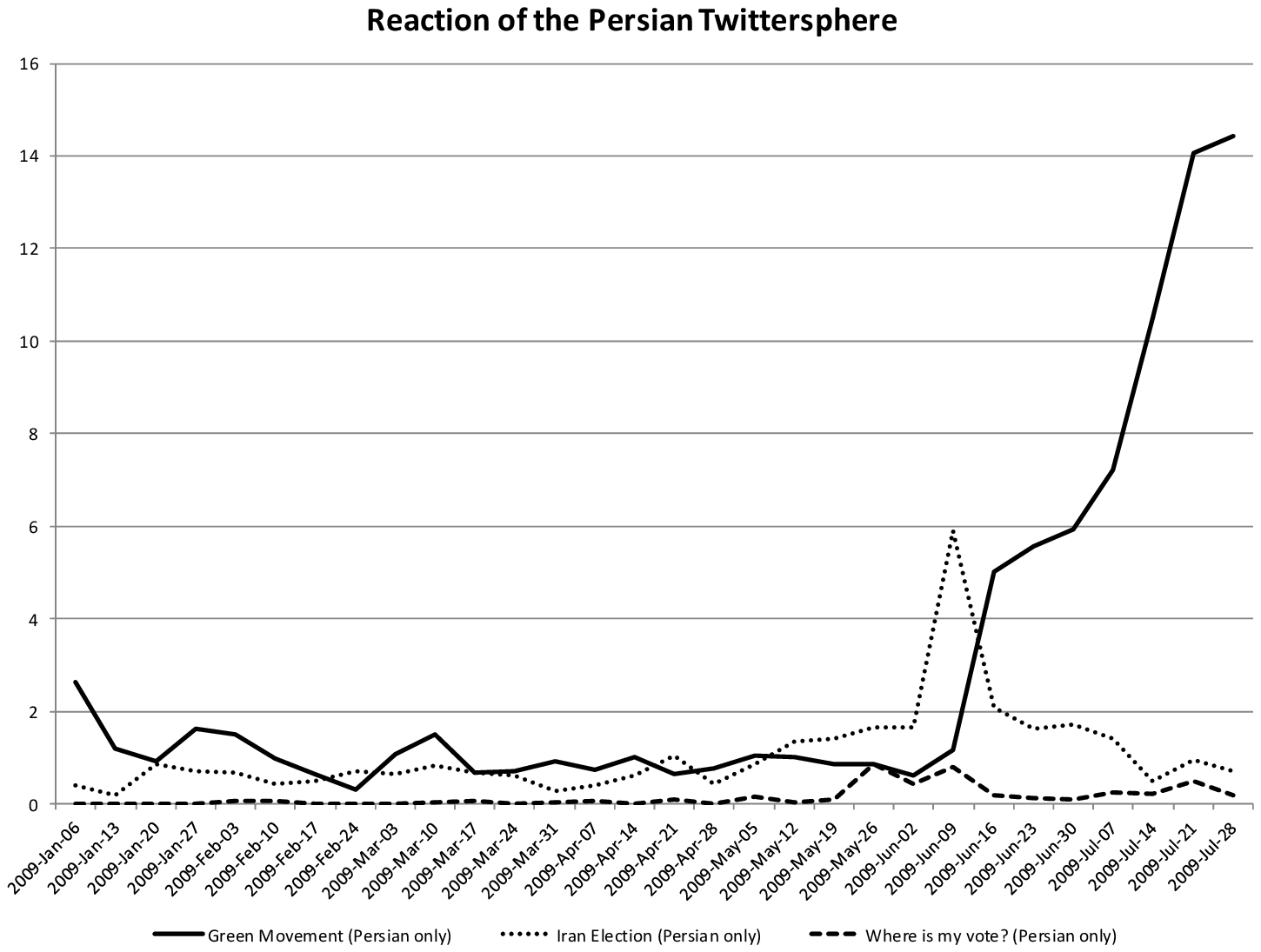}
\par\end{centering}
\caption{\foreignlanguage{english}{Evolution in tweets per-mille of the topics related to the 2009 Iranian
presidential elections in Farsi and Finglish.}}
\end{figure}

Fig. 10 covers the tweets regarding the 2008 US elections; to that
end the queries \emph{``obama OR biden'' }and \emph{``mccain OR
palin'' }were used, obtaining 866,922 and 965,102 tweets, respectively,
for the period going from March 1, 2007 to November 4, 2008. The figure
reveals that conversation about Obama and Biden was ahead of that
about Republicans until the Republican National Convention; after
that, they were behind consistently until the election day. After
that day, tweets about the loosing ticket dropped and tweets about
Obama and Biden went back to campaign levels except for the inauguration
day. 

Please note that even though those tweets correspond to the whole
Twitterverse during those elections I am not making any claim about
the clairvoyance of tweets to forecast electoral results; what is
clear on the other hand is that conversation on Twitter was driven
by the electoral events, if that conversation volume is a proxy to
cast ballots it is a very different issue\footnote{For a post-morten of the 2008 US Election and the feasibility of using
Twitter data for its forecast see \cite{gayo2011don}; for a thorough
analysis on the predictive ability of Twitter regarding elections
see \cite{gayo2013meta}.}.

Figures 11 and 12 cover the events regarding the 2009 Iranian presidential
elections and the protests to took place immediately after them. To
that end the following queries were used\footnote{Both Farsi and Finglish (a romanization of Farsi common in Internet
fora) were used.}: \textFR{\emph{انتخابات}}\emph{ OR ente{*}ab{*}} (elections), \textFR{\emph{جنبش
سبز }}\emph{OR ``rahe sabz''} (green movement), and \textFR{ر\emph{أی
من کجاست}}\emph{ OR ``raye man kojast''} (where is my vote) to cover
the Persian Twittersphere; and their counterparts \emph{“iran election”
OR iranelection}, \emph{“green movement”}, and \emph{“where is my
vote”}, for the global Twittersphere–mostly English-speaking. 

What we found is that use of Twitter in Iran was at that moment mostly
anecdotal; actually, the queries in Farsi and Finglish produce 1,514;
4,597; and 211 results, respectively. Now, compare that with the results
outside the Persian Twittersphere: 1,758,216; 6,045; and 3,589 tweets,
respectively. Fig. 11 reveals that the volume of Iranian tweets is
minimal when compared with the volume of international tweets covering
the elections, not so those covering the ``green movement'' or its
slogan ``where is my vote'' outside Iran. Fig. 12 shows that even
among the Persian Twittersphere the volume of tweets discussing the
elections is also small, and only those related to the ``Green Movement''
experimented a substantial raise after the publication of the results. 

The dataset does not cover the follow up of those protests but with
the data at hand it seems clear that the so-called Twitter Revolution
\cite{sullivan2009revolution} was more an international movement
fueled by the hype of news outlets than an actual reality inside Iran–see
\cite{morozov2009iran} for a thorough analysis on this.

\section*{Mr Dorsey, tear down this paywall!}

To sum up, after collecting the whole production of tweets from March
2006 to July 2009 we may say that (1) tweets closely follow the development
of newsworthy events, and that Twitter-based collective narration
has been taking place from the early phases of the platform; (2) the
peaks and valleys may be a proxy for notoriety but it is difficult
to ascertain other more interesting metrics; and (3) events taking
place in non-English speaking countries may appear in the dataset
but the coverage is sketchy at best. 

Still, I claimed earlier in this paper that Twitter data has been
considered of historical and cultural value, and that full access
is considerably expensive; hence, you may wonder about the cost of
this dataset in the market. To that end I'm using the prices published
by Sifter at the moment of this writing\footnote{http://sifter.texifter.com/\#sifter-pricing}:
\begin{quotation}
\$20 per day of data retrieval

\$30 per 100,000 tweets
\end{quotation}
Given that the archive here described covers 1,217 days and contains
1,483,823,453 tweets it would amount to \$469,487.04! The question
is simple, does this material worth the value? Should public funds
be used to acquire such data time and again to perform different research
projects? My position, of course, is that their actual cost is much
lower and taking into account it is quite dated the price should be
considerably lower; still, the price would depend on the data being
offered as a service or as an appliance.

Granting access to this data is perfectly doable while honoring Twitter's
TOS; indeed, I could offer up to 50,000 tweets per day per user for
free\footnote{See section F.2 of https://dev.twitter.com/overview/terms/policy.}.
However, I'm afraid that for most research questions that would mean
months to obtain the data, only to eventually find that it is not
particularly useful. Actually, most of the value of the data lies
on the possibility of freely exploring it in search of interesting
research opportunities\footnote{Please note that I'm talking about exploratory research and not suggesting
to apply the ``data piñata'' method.}. Providing that in a centralized manner would require a substantial
infrastructure and, still, bulks of more than 50,000 tweets would
not be possible.

Because of that, I dare to suggest that a canon Twitter Archive should
not be offered as a service but instead delivered by postal mail as
a physical item–e.g., a SSD containing an ElasticSearch snapshot.
In that case the cost would be approximately the price of the medium
plus the labor cost to create it plus handling and delivery costs.
Such an item should be available to anyone performing research in
a public or private institution and subject to a non-disclosure agreement.
I firmly believe that such an approach would bring up extremely interesting
research with little burden to Twitter and virtually no risks to the
privacy of their users\footnote{A similar approach was followed by Microsoft Research to released
a query log; teams were sent a DVD after signing a NDA (Non-disclosure
agreement); the program has been discontinued but it is probably because
of the data being dated and not due to privacy concerns.}. I also think that Twitter Inc. could test this idea rather easily
in a second installment of the Twitter Data Grants and provide access
to a much larger number of teams.

\section*{Conclusions}

All public tweets published between March 2006 and July 2009 (almost
1.5 billion) were collected using the available Twitter API; the dataset
is relatively large (about 1 TB when decompressed) but can be reduced
to a more manageable form by severely reducing the stored metadata.
That way it is possible to use ElasticSearch to provide a searchable
index fitting in an relatively inexpensive workstation. Performance
is reasonable if using a SSD and, while not real-time, average queries
can be issued in an interactive fashion. 

The archive is interesting because of its historical value given that
the best known Twitter behaviors and phenomenons such as hashtags,
trending topics, retweets or follow Fridays appear in the archive.
Moreover, some historically important events such as the US 2008 Presidential
Elections or the so-called Green Revolution of Iran are also covered
by the data. Still, the coverage of such events is highly biased since,
at those moments, Twitter was far from being a global phenomenon. 

Anyway, I consider that this dataset offers many opportunities for
researchers and it can be replicated following the directions provided
the paper. It must be noted, however, that anyone doing that and respecting
Twitter TOS would require a large amount of time to complete it–years,
actually. Moreover, if a relevant number of researchers decided to
replicate this effort it would mean a waste of resources both on the
part of the researchers and on the part of Twitter Inc. That's why
I suggest that Twitter itself could market the archive as a physical
item providing it to affiliated researchers and subject to a non-disclosure
agreement.

\section*{Acknowledgments}

I would like to thank David J. Brenes and David Moreno-García who
were brave enough to try to fight the original raw dataset by other
means. I want also to thank Darío Álvarez-Gutiérrez who provided invaluable
help with the Neo4J model in the appendix.

\bibliographystyle{plain}
\addcontentsline{toc}{section}{\refname}\bibliography{Twitter-Historical-Archive-paper}

\begin{thebibliography}{10}

\bibitem{allen2013update}
Erin Allen.
\newblock Update on the twitter archive at the library of congress.
\newblock In {\em Library of Congress Blog}, volume~4, 2013.

\bibitem{bruns2016twitter}
Axel Bruns and Katrin Weller.
\newblock Twitter as a first draft of the present: and the challenges of
  preserving it for the future.
\newblock In {\em Proceedings of the 8th ACM Conference on Web Science}, pages
  183--189. ACM, 2016.

\bibitem{burgess2013promised}
Jean Burguess.
\newblock As promised, here is the link to oldtweets, a searchable archive of
  the first year of twitter http://kellan.io/oldtweets \#compdata13.
\newblock In {\em Twitter}, 2013.

\bibitem{cha2010measuring}
Meeyoung Cha, Hamed Haddadi, Fabricio Benevenuto, and P~Krishna Gummadi.
\newblock Measuring user influence in twitter: The million follower fallacy.
\newblock {\em ICWSM}, 10(10-17):30, 2010.

\bibitem{douglas2007twitter}
Nick Douglas.
\newblock Twitter blows up at sxsw conference.
\newblock In {\em Gawker.com}, 2007.

\bibitem{gayo2011don}
Daniel Gayo-Avello.
\newblock Don't turn social media into another'literary digest'poll.
\newblock {\em Communications of the ACM}, 54(10):121--128, 2011.

\bibitem{gayo2013meta}
Daniel Gayo-Avello.
\newblock A meta-analysis of state-of-the-art electoral prediction from twitter
  data.
\newblock {\em Social Science Computer Review}, page 0894439313493979, 2013.

\bibitem{king2010announcing}
Ryan King.
\newblock Announcing snowflake.
\newblock {\em Twitter Engineering Blog}, 2010.

\bibitem{krikorian2014introducing}
Raffi Krikorian.
\newblock Introducing twitter data grants.
\newblock {\em Twitter Engineering Blog}, 2014.

\bibitem{krikorian2014twitter}
Raffi Krikorian.
\newblock Twitter \#datagrants selections.
\newblock {\em Twitter Engineering Blog}, 2014.

\bibitem{kwak2010twitter}
Haewoon Kwak, Changhyun Lee, Hosung Park, and Sue Moon.
\newblock What is twitter, a social network or a news media?
\newblock In {\em Proceedings of the 19th international conference on World
  wide web}, pages 591--600. ACM, 2010.

\bibitem{mccreadie2012building}
Richard McCreadie, Ian Soboroff, Jimmy Lin, Craig Macdonald, Iadh Ounis, and
  Dean McCullough.
\newblock On building a reusable twitter corpus.
\newblock In {\em Proceedings of the 35th international ACM SIGIR conference on
  Research and development in information retrieval}, pages 1113--1114. ACM,
  2012.

\bibitem{mcgill2016can}
Andrew McGill.
\newblock Can twitter fit inside the library of congress?
\newblock {\em The Atlantic}, 2016.

\bibitem{morozov2009iran}
Evgeny Morozov.
\newblock Iran: Downside to the" twitter revolution".
\newblock {\em Dissent}, 56(4):10--14, 2009.

\bibitem{morstatter2014biased}
Fred Morstatter, J{\"u}rgen Pfeffer, and Huan Liu.
\newblock When is it biased?: assessing the representativeness of twitter's
  streaming api.
\newblock In {\em Proceedings of the 23rd International Conference on World
  Wide Web}, pages 555--556. ACM, 2014.

\bibitem{parr2010twitter}
Ben Parr.
\newblock Twitter improves trending topic algorithm: Bye bye, bieber!
\newblock In {\em Mashable}, 2010.

\bibitem{raymond2010tweet}
Matt Raymond.
\newblock How tweet it is!: Library acquires entire twitter archive.
\newblock In {\em Library of Congress blog}, volume~14, 2010.

\bibitem{rogers2013debanalizing}
Richard Rogers.
\newblock Debanalizing twitter: The transformation of an object of study.
\newblock In {\em Proceedings of the 5th Annual ACM Web Science Conference},
  pages 356--365. ACM, 2013.

\bibitem{scola2015library}
Nancy Scola.
\newblock Library of congress' twitter archive is a huge \#fail.
\newblock {\em Politico}, 2015.

\bibitem{singletary2010upcoming}
Taylor Singletary.
\newblock Upcoming changes to the way status ids are sequenced.
\newblock {\em Twitter Development Talk}, 2010.

\bibitem{sullivan2009revolution}
Andrew Sullivan.
\newblock The revolution will be twittered.
\newblock {\em The Atlantic}, 13, 2009.

\bibitem{tufekci2014big}
Zeynep Tufekci.
\newblock Big questions for social media big data: Representativeness, validity
  and other methodological pitfalls.
\newblock In {\em Eighth International AAAI Conference on Weblogs and Social
  Media}, 2014.

\bibitem{watts2013computational}
Duncan~J Watts.
\newblock Computational social science: Exciting progress and future
  directions.
\newblock {\em The Bridge on Frontiers of Engineering}, 43(4):5--10, 2013.

\bibitem{zhuang2014building}
Yi~Zhuang.
\newblock Building a complete tweet index.
\newblock {\em Twitter Engineering Blog}, 2014.

\bibitem{zimmer2015twitter}
Michael Zimmer.
\newblock The twitter archive at the library of congress: Challenges for
  information practice and information policy.
\newblock {\em First Monday}, 20(7), 2015.

\end{thebibliography}

\pagebreak{}

\section*{Appendix 1: Octave code to produce the tweet ids of the archive}

The following Octave code produces tweet ids according to the different
generation approaches which were detected in the dataset. This approach
produces 2,292,166,175 ids and, thus, about 36\% of these ids correspond
to non-public, deleted or simply nonexistent tweets; I offer it for
those not willing to download a multi-gigabyte file full of ids. Those
with a more paranoid nature can used the following code instead \emph{1:3061014649}
although it's going to take much more time to complete.

\begin{multicols}{2} 
\begin{verbatim}
20:81803 
81803:10:5317478 
5317478:5951471 
5951471:10:33659941 
33659941:34051542 
34051542:10:749778882 
749778882:797700951 
797700951:10:798082536 
798082536:861278101 
861278101:10:861796399 
861796399:907582571 
907582571:10:907936108 
907936108:10:908894500 
908894500:920578209 
920578209:10:920903970 
920903970:948996649 
948996649:10:950233829 
950233829:957352345 
957352345:10:957603791 
957603791:989085799 
989085799:10:989696020 
989696020:1062054690 
1062054690:10:1062633411 
1062633411:1063043430 
1063043430:10:1067177961 
1067177961:1268484169 
1268484169:10:1268752486 
1268752486:1276604442 
1276604442:10:1278491542 
1278491542:1305643870 
1305643870:10:1308469567 
1308469567:1337207851 
1337207851:10:1337561777 
1337561777:1341303857 
1341303857:10:1341654616 
1341654616:1347134019 
1347134019:10:1347350683 
1347350683:1358197730 
1358197730:10:1358777850 
1358777850:1365719225 
1365719225:10:1365920715 
1365920715:1433622936 
1433622936:10:1434276794 
1434276794:1445739899 
1445739899:10:1445939272 
1445939272:1467920729 
1467920729:10:1469118986 
1469118986:1469829667 
1469829667:10:1470202281 
1470202281:1476157039 
1476157039:10:1477122821 
1477122821:1489587493 
1489587493:10:1490448333 
1490448333:1494130684 
1494130684:10:1496083713 
1496083713:1682400687 
1682400687:10:1690038860 
1690038860:1711490733 
1711490733:10:1711821385 
1711821385:1726965818 
1726965818:10:1727193439 
1727193439:1734319873 
1734319873:10:1735002265 
1735002265:1986606277 
1986606277:10:1986848681 
1986848681:2023225505 
2023225505:10:2023747452 
2023747452:2048511605 
2048511605:10:2051073819 
2051073819:2111076679 
2111076679:10:2113946682 
2113946682:2130679870 
2130679870:10:2136747540 
2136747540:2202236683 
2202236683:10:2202553995 
2202553995:2313545593 
2313545593:10:2322204712 
2322204712:2408271108 
2408271108:10:2416149025 
2416149025:2453681374 
2453681374:10:2458487491 
2458487491:2486851754 
2486851754:10:2490430312 
2490430312:2530881466 
2530881466:10:2548066056 
2548066056:2831755333 
2831755333:10:2851555775 
2851555775:3061014649
\end{verbatim}
\end{multicols}

\pagebreak{}

\section*{Appendix 2: Exercise for the reader}

The interested reader could try to create a Neo4J implementation of
the archive using the following model. Please note that to include
the :FOLLOWS and :IS\_FOLLOWED relations you will need to download
the Twitter User Graph compiled by KAIST \cite{kwak2010twitter}\footnote{https://an.kaist.ac.kr/traces/WWW2010.html}.

\begin{figure}[H]
\includegraphics[width=12cm]{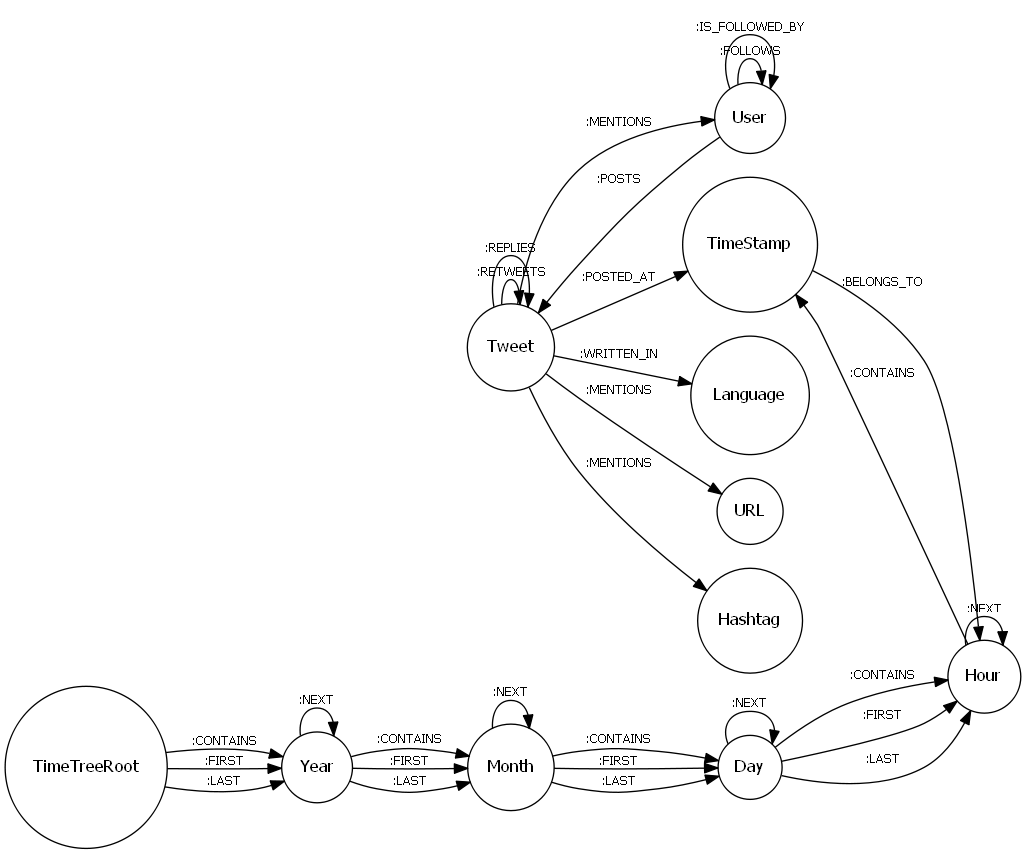}\caption{\foreignlanguage{english}{A tentative model for a Neo4J version of the Twitter Historical Archive
combined with the Twitter User Graph.}}
\end{figure}
\selectlanguage{english}%

\end{document}